\documentclass{optica-article}

\journal{opticajournal} 

\articletype{Research Article}

\usepackage{lineno}

\begin{document}

\title{Dynamic manipulation of graphene plasmonic skyrmions}

\author{Ni Zhang,\authormark{1,2} Xinrui Lei,\authormark{1,2} Jiachen Liu,\authormark{1,2} and Qiwen Zhan\authormark{1,2,*}}

\address{\authormark{1}School of Optical-Electrical and Computer Engineering, University of Shanghai for Science and Technology, Shanghai 200093, China\\
\authormark{2}Zhangjiang Laboratory, Chinese Academy of Sciences (CAS), 100 Haike Road, Shanghai 201204, China\\
}

\email{\authormark{*}qwzhan@usst.edu.cn} 


\begin{abstract*} 
With the characteristics of ultrasmall, ultrafast and topological protection, optical skyrmions has great prospects in application of high intensity data stroage, high resolution microscopic imaging and polarization sensing. The flexible control of the optical skyrmions is the premise of practical application. At present, the manipulation of optical skyrmions usually relies upon the change of spatial structure, which results in a limited-tuning range and a discontinuous control in the parameter space. Here, we propose continuous manipulation of the graphene plasmons skyrmions based on the electrotunable properties of graphene. By changing the Fermi energy of one pair of the standing waves and the phase of the incident light can achieve the transformation of the topological state of the graphene plasmons skyrmions, which can be illustrated by the change of the skyrmion number from 1 to 0.5. The direc manipulation of the graphene plasmons skyrmions is demonstrated by the simulation results based on the finite element method. Our work suggests a feasible way to flexibly control the optical skyrmions topological field, which can be used for novel integrated photonics devices in the future.

\end{abstract*}

\section{Introduction}
Skyrmions was proposed by British physicist Tony Skyrme in the 1960s~\cite{skyrme1962unified1}, which is a particle-like spin texture at micro-nano scale and with the properities of topological protection~\cite{yu2010real2, nagaosa2013topological3}. Recently it has been introducted to the realm of optics~\cite{lei2021photonic42, lei2023metastability43, gutierrez2021optical45, gao2020paraxial46, lin2021microcavity47, RNshen48}. Surface plasmon polaritons (SPPs) have been widely employed to generate skyrmionic textures since the two-dimensional confinement of light in evanescent field provides a smooth domain for topological textures, which can be constructed by either electric/magnetic field or spin angular momentum. The versatile topogical textures provides new degrees of freedom for shaping vectorial fields and encoding information, which has broad application prospects in ultrafast vector imaging~\cite{lei2023metastability43}, nanoscale metrology~\cite{yang2023spin44}, topological hall devices~\cite{karnieli2021emulating24}, etc. Therefore, it is of great significance to realize flexible manipulation of optical skyrmions.

Currently, the manipulation of optical skyrmions is mostly realized by designing different spatial structure~\cite{deng2022observation5, ghosh2023spin6}, which results in a limited-tuning range, a complex operating system, making it diffcult to realize devices integration. And once the structure is fixed, only a specific topological state can be realized~\cite{dai2022ultrafast25, guo2020meron26, deng2022observation27}: a direct transformation from one topological state to another lacks concrete transition process, which is not conducive to a clear understanding of the topological properties of optical skyrmions and their application in practice.  Graphene plasmons (GPs)~\cite{jablan2009plasmonics28, minovkoppensich2011graphene9}, as a kind of surface wave, is essentially the collective oscillations of free Dirac electrons in graphene coupling to electromagnetic fields, have attracted a great deal of attention owing to their strong field confinement, short wave-lengths and continuous electrical tunability~\cite{wunsch2006dynamical7, fei2012gate8, chen2012optical10}. GPs can not only be regulated by the nanostructure deposited on the graphene, but also impressed voltage based on the material properities, which is not available in other metal surface plasmons~\cite{chan2008localized29, knight2012aluminum30, langhammer2008localized31}. 

In this paper, we constructed GPs skyrmions based on the interaction between vector light field and graphene hexgonal structure. Dynamic regulation of optical skyrmions by both the material property and vector field was proposed. Numerical simulations are performed to verify the proposed technique. In section 2, the principle of GPs skyrmions and topological transformation by continuity regulation by changing the vector field and the Fermi energy of graphene is discussed. In section 3, the theoretical analysis and full-wave electromagnetic simulations were carried out to demonstrate the feasibility of dynamic regulation of GPs skyrmions based on vector field and the electrical tunable properties of graphene. Finally, the main conclusions are summarized in section 4. Our work opens an avenue for engineering the optical skyrmions topological states continuously and is promising for application in integrated potoelectronic devices, including optical sensors and light modulators. 

\section{Working principle}

\begin{figure}[ht!]
\centering\includegraphics[width=0.8\textwidth]{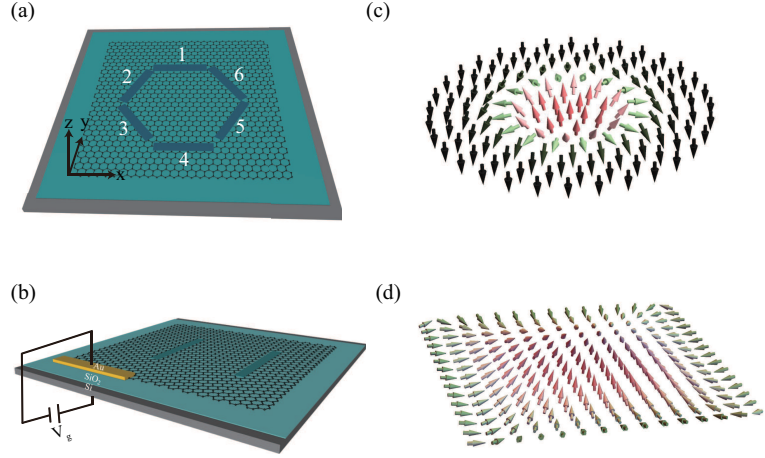}
\caption{Schematic diagram of dynamic manipulation of optical skyrmions by changing the Fermi energy of graphene by gate-voltages. (a) Graphene structure consists of three pairs of parallel slits with azimuths of -$\pi$/3, 0 and $\pi$/3. (b) Back-gate tunable the Fermi energy of a pair of parallel graphene slit. (c-d) GPs skyrmions represented by the local electric fields generated by the interference of standing graphene plasmons, where the topology can be manipulated by changing the Fermi enenrgy of the graphene slits.}
\label{fig1}
\end{figure}

The schematic diagram of continuous manipulation of optical skyrmions based on the electrical tunable properities of graphene is shown in Fig.~\ref{fig1}(a) and Fig.~\ref{fig1}(b). Six gratings creating a hexagon are deposited on the graphene layer which is placed on the Si substrate with a 300~nm SiO$_{2}$. Each pair of the parallel slits create a graphene plasmons standing wave that travels along the surface of the graphene. Therefore , the field at the center of the hexagon is the superposition of three pairs of the graphene plasmons standing waves passing through the slits. Changing the Fermi energy of graphene by applying gate-voltages, the graphene plasmons stangding waves can be regulated as shown in Fig.~\ref{fig1}(b). The z-component of the GPs along the $x-y$ plane can be expressed as: 
\begin{equation}
E_{z}^{(\omega)}=E_{0}e^{-{\left|{k_{z}z}\right|}}{\cos\{k_{GP_{s}}[\cos{(\theta)x}+\sin{(\theta)}y]\}}
\label{eqn1}
\end{equation}
Therefore, the field in the center of the hexagon graphene slits can be regarded as the superposition of standing waves with a azimuth angle of -$\frac{\pi}{3}$, $0$, $\frac{\pi}{3}$ , respectively, which are formed by three pairs of parallel graphene slits numbered by 14, 25 and 36 in Fig.~\ref{fig1}(a) . The $z$ component of electric field at the center can be expressed as~\cite{tsesses2018optical11}:
\begin{equation}
E_{z}^{(\omega)}=E_{0}e^{-{\left|{k_{z}z}\right|}}{\sum_{\theta=-{\frac{\pi}{3}},{0},{\frac{\pi}{3}}}}{\cos\{k_{GP_{s}}[\cos{(\theta)x}+\sin{(\theta)}y]\}} 
\label{eqn2}
\end{equation}
where $E_{0}$ is a real normalization constant; $k_{GPs}$ is the transverse components of the wave vector of the graphene plasmons; $k_{z}$ is the axial component of the wave vector. And ${k_{GPs}}^{2}-{k_{z}}^{2}={k_{0}}^{2}$, $k_{0}$ is the free-space wave number.

The transverse electric field components satisfy the following form, which can be derived from Maxwell's equations:
\begin{equation}
\left[
\begin{array}{cc}
E_{x}^{(\omega)} \\
E_{y}^{(\omega)} \\
\end{array}
\right]=-E_{0}\frac{\left|{k_{z}}\right|}{k_{GPs}}e^{-\left|{k_{z}}\right|z}{\sum_{\theta=-\frac{\pi}{3},0,\frac{\pi}{3}}}
\left[
\begin{array}{cc}
\cos(\theta) \\
\sin(\theta) \\
\end{array}
\right]
\sin\{k_{GPs}[\cos(\theta)x+\sin(\theta)y]\}
\label{eqn3}
\end{equation}
It is well known that, the topological properties of a skyrmionic configuration can be characterized by the skyrmion number (topological invariant) $S$, which can be defined as~\cite{tsesses2018optical11, davis2020ultrafast16}:
\begin{equation}
S=\frac{1}{4\pi}\int_{A}\left({\vec{e}\cdot{\bigg [}{\frac{\partial {\vec{e}}}{\partial x}\times\frac{\partial {\vec{e}}}{\partial y}}}{\bigg ]}\right)dA
\label{eqn8}
\end{equation}
\begin{equation}
\vec{e}=\frac{{Re\left\{(E_{x}^{(\omega)},E_{y}^{(\omega)},E_{z}^{(\omega)})\right\}}}{\sqrt{{\left |E_{x}^{(\omega)}\right |^{2}}+{\left |E_{y}^{(\omega)}\right |^{2}}+{\left |E_{z}^{(\omega)}\right |^{2}}}}
\label{eqn9}
\end{equation}

Here, $\vec{e}$ is a real, normalized, three-component field; the area A covers one unit cell of the lattice; $x$ and $y$ are directions in the 2D plane. The skyrmion number $S$ is robust to deformations of the field $\vec{e}$ as long as $\vec{e}$ remains nonsingular and maintains the periodicity of the lattice.

From Eq. (\ref{eqn2}) and Eq. (\ref{eqn3}), it can be known that the intensity of the GPs standing wave at the center of the hexagonal slits can be controlled by changing the wavevetor $k_{GPs}$ of the GPs. The dispersion relation of GPs can be obtained by calculating the intrinsic vibration solution of passive electromagnetic wave at the interface of graphene. The expression satisfies the following form:
\begin{equation}
\frac{\varepsilon_{1}}{\sqrt{\varepsilon_{1}{k_{0}}^{2}-{k_{GPs}}^{2}}}+\frac{\varepsilon_{2}}{\sqrt{\varepsilon_{2}{k_{0}}^{2}-{k_{GPs}}^{2}}}=-\frac{i\sigma}{\omega\varepsilon_{0}}
\label{eqn4}
\end{equation}
Here, $\varepsilon_{1}$, $\varepsilon_{2}$ is the dielectric constant of the up and down interface medium of graphene, $k_{0}$ is the wave vector in vacuum and $\sigma$ is the conductivity of graphene, which satisfies~\cite{nikitin2011fields12}:
\begin{gather}
\sigma_{intra}=\frac{2e^{2}k_{B}T}{\pi{\hbar}^{2}} \frac{i}{\omega+i{\tau}^{-1}} \ln[2\cosh(\frac{E_{F}}{2k_{B}T})]   \notag\\
 \sigma_{inter}=\frac{e^{2}}{4\hbar} [\frac{1}{2}+\frac{1}{\pi}\arctan(\frac{\hbar\omega-2E_{F}}{2k_{B}T})-\frac{i}{2\pi}\ln{\frac{(\hbar\omega+2E_{F})^{2}}{(\hbar\omega-2E_{F})^{2}+(2k_{B}T)^{2}}}]       \notag\\
\sigma=\sigma_{intra}+\sigma_{inter}    
\label{eqn5}
\end{gather}
Therefore, when the excitation beams are fixed, the wave vector of the GPs is related to the conductivity of graphene, which is the function of the Fermi energy. The Fermi energy $E_{F}$ in graphene with respect to the Dirac point can be easily tuned by changing the carrier density $n_{s}$ with $E_{F}=\hbar{\nu_{F}}\sqrt{\pi{n_{s}}}$~\cite{RN13, RN14}, proving an efficient method for active control GPs. Where $\nu_{F}=1\times{10}^{6}$m/s is the Fermi speed, $n_{s}$ is the carrier concentration. Figure~\ref{fig1}(b) illustrates the back-gate configuration, where the voltage is applied between the Si substrate and the Au electrode connected to graphene layer. The Au electrode is placed far away from the graphene slits structures to avoid the influence on the GPs. In this way, the wave vector of each GPs standing wave can be changed and then controled the superposition field distribution. Fig.~\ref{fig1}(c) and Fig.~\ref{fig1}(d) is the vector representation of the local unit vector of the electric field before and after the Fermi energy changes in one of the three pairs of graphene slits, showing that a transition of skyrmion topology and the arrows indicate the direction of the local unit vector of the electric field.

On the other hand, when the Fermi energy of the three pairs of graphene slits is fixed, the phase difference of each pair of the standing wave can be control by changing the phase and polarization of the excitation beams of the GPs wave~\cite{RN13, RN15}. Considering the phase difference of the standing wave, the expresstion of the electric field distribution at the center of the graphene hexagonal slits can be described as following:
 \begin{equation}
E_{z}^{(\omega)}=E_{0}e^{-{\left|{k_{z}z}\right|}}{\sum_{\theta=-{\frac{\pi}{3}},{0},{\frac{\pi}{3}}}}{e^{i\phi_{\theta}}}{\cos\{k_{GP_{s}}[\cos{(\theta)x}+\sin{(\theta)}y]\}} 
\label{eqn6}
\end{equation}
\begin{equation}
\left[
\begin{array}{cc}
E_{x}^{(\omega)} \\
E_{y}^{(\omega)} \\
\end{array}
\right]=-E_{0}\frac{\left|{k_{z}}\right|}{k_{GPs}}e^{-\left|{k_{z}}\right|z}{\sum_{\theta=-\frac{\pi}{3},0,\frac{\pi}{3}}}{e^{i\phi_{\theta}}}
\left[
\begin{array}{cc}
\cos(\theta) \\
\sin(\theta) \\
\end{array}
\right]
\sin\{k_{GPs}[\cos(\theta)x+\sin(\theta)y]\}
\label{eqn7}
\end{equation}
 
Here, $\phi_{\theta}$ is the phase difference of the GPs stangding wave formed by a pair of parallel slits with the azimuth of $\theta$. In order to obtain the maximum coupling effect, the polarization of the excitation beams need to perpendicular to the slits. When a pair of the parallel graphene slits are excited by two exications beam with the phase of 0 and the opposite polarization, the slits act as the reflection boundaries and the plasmons wave will propagating along opposite direction with the phase diffeerence of $\pi$. Therefore, the GPs stangding wave can be formed by two separate parallel graphene slits, which interference with each other in the middle region. So, the intensity at the center of the GPs standing wave depends on phase difference of two GPs waves, which can be controlled by changing the phase and polarization of the excitation beams. In experimental, the phase difference of the graphene standing wave can be obtained by shifting the distance from the excitation wave to the graphene slits. 

\section{Simulation and discussion}

\begin{figure}[!htbp]
\centerline{\includegraphics[width=0.8\textwidth]{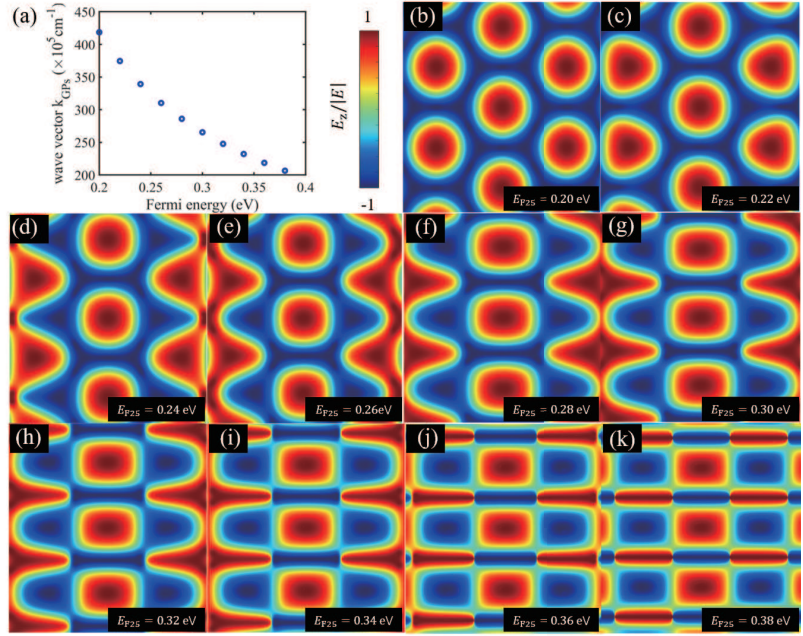}}
\caption{Calculated electric field distribution of GPs skyrmions as changes the Fermi energy of one pair of the GPs standing wave. (a) The dispersion relation of the GPs as $E_{F14}$=0.20$\sim$0.38~eV. And $\theta_{14}$=$\theta_{25}$ =$\theta_{36}$=0~eV, $E_{F25}$=$E_{F36}$=0~eV. The blue circles represent the data of the simulation result. (b)-(k) Axial electric field distribution at different Fermi energy, according to Eq. (\ref{eqn2}). The color scale indicates the value of $E_{z}{/}\left | E \right |$.}
\label{fig2}
\end{figure}

For convenience, the Fermi energy and the phase difference of the three standing waves, which formed by the parallel graphene slits in Fig.~\ref{fig1}(a) are represented by $E_{F14}$, $E_{F25}$, $E_{F36}$, $\theta_{14}$, $\theta_{25}$ and $\theta_{36}$, respectively. Considering that the six excitation beams have the same phase and polarization, that is, $\theta_{14}$=$\theta_{25}$=$\theta_{36}$=0, we calculated the change of the optical skyrmions electric field at the center of the graphene hexagon when the Fermi energy of only one of the three standing waves is changed. Based on Eq. (\ref{eqn4}) and Eq. (\ref{eqn5}), the dispersion diagram of the GPs wave as the changes of the Fermi energy is shown in Fig.~\ref{fig2}(a) and the wavelength of excitation light is $\lambda_{0}$=10.653~$\mu$m. It can be known that, with the increases of the Fermi energy, the real part of the wave vectors of the GPs decreases gradually and the wave vector of $E_{F}$=0.40~eV is approximately half of that of $E_{F}$=0.20~eV shown as the blue circle in Fig.~\ref{fig2}(a). Substituting the wave vectors of different Fermi energy into Eq. (\ref{eqn2}) and Eq. (\ref{eqn3}), the normalized results of the axial (out-of-plane) electric field were obtained by numerical simulation as shown in Fig.~\ref{fig2}(b)-(k). Here, $E_{F25}$=$E_{F36}$=0 and $E_{F14}$ is continuously changed from 0.20~eV to 0.38~eV with a step of 0.02~eV. The color scale indicates the value of $E_{z}^{(\omega)}/\sqrt{(E_{x}^{(\omega)})^{2}+(E_{y}^{(\omega)})^{2}+(E_{z}^{(\omega)})^{2}}$. 

The axial electric field has the form of a hexagonally symmetric lattice as the standing wave of the three pairs of parallel slits have the same Fermi energy $E_{F14}$=$E_{F25}$=$E_{F36}$=0.20~eV as shown in Fig.~\ref{fig2}(b). Keeping $E_{F25}$=$E_{F36}$=0.20~eV unchanged and gradually increases $E_{F14}$ with a step of 0.02~eV, the shape of the optical skyrmions at the center along the vertical direction changes gradually from circle to square as shown in Fig.~\ref{fig2}(c)-(k). At the same time, the shape of the optical skyrmions in the left and right areas along the vertical direction changes from circle to equilateral triangle and then the base of the equilateral triangles shrinks gragually to an isosceles triangle until it becomes a rectangle. 
To further illustrate the change of the topological states of GPs skyrmions, the skyrmions number $S$ of each site was calculated based on Eq. (\ref{eqn8}) and Eq. (\ref{eqn9}). For the axial electric field has the form of a hexagonally symmetric lattice with skyrmions number $S$ of 0.9934 ($\approx$1)corresponding to $E_{F14}$=$E_{F25}$=$E_{F36}$=0.20~eV, while for that of the axial electric field which has the form of a rectangle lattice, the skyrmions number $S$ of 0.5097 ($\approx$0.5) corresponding to $E_{F14}$=0.38~eV, $E_{F25}$=$E_{F36}$=0.20~eV. This shows that the continuous manipulation of optical skyrmions topological states can be achieved by changing the Fermi energy of GPs. 

On the other hand, the Fermi energy of the three GPs standing wave keep the same, here, $E_{F14}$=$E_{F14}$=$E_{F14}$=0.20~eV, the optical skyrmions electric field at the center of the graphene hexagon was calculated when changed the phase difference of one of the three standing wave. Supposing that the excitation waves of the two pairs of the graphene slits with azimuths of -$\pi$/3 and $\pi$/3 have the same phase and polarization, that means $\theta_{25}=\theta_{36}$=0, while the other pair of graphene slit with azimuth of 0 has the same polarization but the phase difference of $\pi$. Therefore, the phase diference of the graphene hexagon is $\theta_{14}=\pi$. The electric field distribution at the center of the graphene hexagon structure satisfies Eq. (\ref{eqn6}) and Eq. (\ref{eqn7}). Therefore, the dynamic manipulation of the optical skyrmions electric field can also be achieved by changing the phase of the incident vector light. 

\begin{figure}[!htbp]
\centerline{\includegraphics[width=0.9\textwidth]{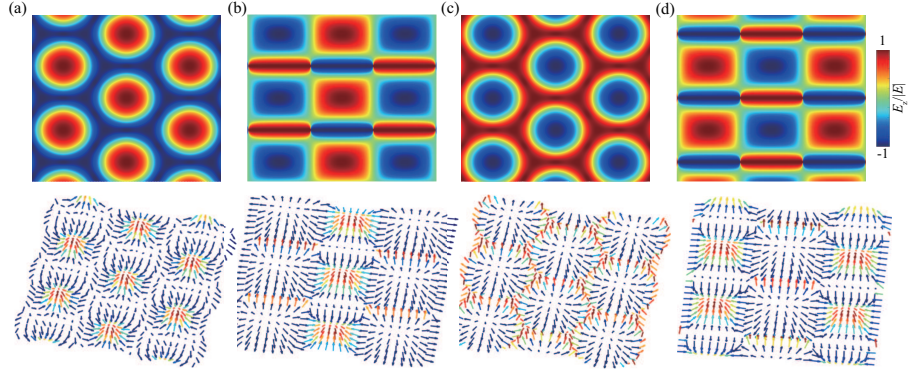}}
\caption{Calculated electric field distribution of the optical skyrmion lattice when changing the Fermi energy $E_{F14}$ and the phase $\phi_{14}$, while $E_{F25}=E_{F25}=$0.20~eV, $\phi_{25}=\phi_{36}=0$ keeping unchanged. (a-d) Normalized $z$ component of electric field, according to Eq. (\ref{eqn2}) and Eq. (\ref{eqn3}) when (a) $E_{F14}=$0.20~eV, $\phi_{14}=0$; (b) $E_{F14}=$0.38~eV, $\phi_{14}=0$; (c) $E_{F14}=$0.20~eV, $\phi_{14}=\pi$; (d) $E_{F14}=$0.38~eV, $\phi_{14}=\pi$, respectively. The vector distributions of the electric field correspond to the intensity distribution in the bottom of (a-d) respectively. }
\label{fig3}
\end{figure}

The vector fields and the intensity distribution at the center of the graphene hexagon slits when changes the Fermi energy of graphene slits or the phase difference of the incident vector light field as shown in Fig.~\ref{fig3}. It is clear that symmetrical hexagon optical skyrmion appears in the electric field corresponding to $E_{F14}=E_{F25}=E_{F36}$=0.20~eV, $\theta_{14}=\theta_{25}=\theta_{36}$=0 of the three GPs standing waves as shown in Fig.~\ref{fig3}(a) with the three-dimensional electric field vector distribution at the bottom forming a N$\acute{\text{e}}$el-type skyrmion~\cite{zhang2021bloch18, tian2023n51}. The direction of the vector field is from the outside to the inside and from bottom to top which indicated by the arrows. The length and the color of arrows indicate the field intensity. Then, only changed the Fermi energy of the GPs standing wave with azimuthal angle of 0 to 0.38~eV, while others remains unchanged comparaed to Fig. \ref{fig3}(a). Hence, $E_{F14}$=0.38~eV, $E_{F25}=E_{F36}$=0.20~eV, $\theta_{14}=\theta_{25}=\theta_{36}$=0 and the shape of the optical skyrmions changes from circle to square shown in Fig. \ref{fig3}(b) with the three-dimensional electric field vector distribution at the bottom forming a N$\acute{e}$el-type skyrmion. 

Using Eq. (\ref{eqn8}) and Eq. (\ref{eqn9}), the calculated skyrmion number of each unit cell's of Fig. \ref{fig3}(a) and Fig. \ref{fig3}(b) is $S\approx$1 and $S\approx$0.5, respectively. And it further illustrated that the continously change of the optical skyrmions topological states can be achieved based on the electrotunable properities of GPs waves. Similarly, the vector field and the intensity distribution at the center of the graphene hexagon slits was also calculated when only changed the phase of the incident vector light field while the Fermi energy of the three GPs standing waves was fixed at 0.2~eV. Changing the phase difference of $\theta_{14}$ from 0 to $\pi$, the axial electric field distribution of the optical skyrmions at the center of the graphene hexagon was shown in Fig. \ref{fig3}(c) with the three-dimensional electric field vector distribution at the bottom form a N$\acute{\text{e}}$el-type skyrmion, where $\theta_{25}=\theta_{36}$=0 and $E_{F14}=E_{F25}=E_{F36}$=0.20~eV. Compared with Fig. \ref{fig3}(a), the added phase difference of the GPs standing wave not only caused the move of the optical skyrmions along the vertical direction, but also the the direction of the electric field vector distribution become opposite as shown at the bottom of Fig. \ref{fig3}(c). Using the same method as above, the calculated skyrmion number of each unit cell's of Fig. \ref{fig3}(c) is $S=-1.01$ ($\approx$-1). It proved that it is feasible to regulate the topological states of GPs skyrmions by changing the phase difference of the incident vector light.

In addition, we further calculated the vector fields distribution and the axial electric field of the optical skyrmions at the center of the graphene hexagon slits when simultaneously changed the Fermi energy and the phase difference of one of the GPs standing waves. The optical skyrmions electric field of the graphene hexagon slits when $E_{F14}$=0.38~eV, $E_{F25}=E_{F36}$=0.20~eV, $\phi_{14}=\pi$, $\phi_{25}=\phi_{36}=0$ is shown in Fig. \ref{fig3}(d). Comparaed to Fig. \ref{fig3}(c), Fig. \ref{fig3}(d) added the phase difference of $\pi$ of the GPs standing wave at azimuth of 0. The shape of the optical skyrmions is similar to that in Fig. \ref{fig3}(b), but the position of a skyrmion in each unit cell of a lattice had shifted, besides, the direction of the vector distribution of the electric field was opposited. And the calculated skyrmion number of each site of Fig. \ref{fig3}(d) is $S=-0.5076$ ($\approx$-0.5). The change of the skyrmion number shows that the topological states of GPs skyrmions can be changed by changing the Fermi energy of the GPs and the incident vector field.

\begin{figure}[!htbp]
\centerline{\includegraphics[width=0.8\textwidth]{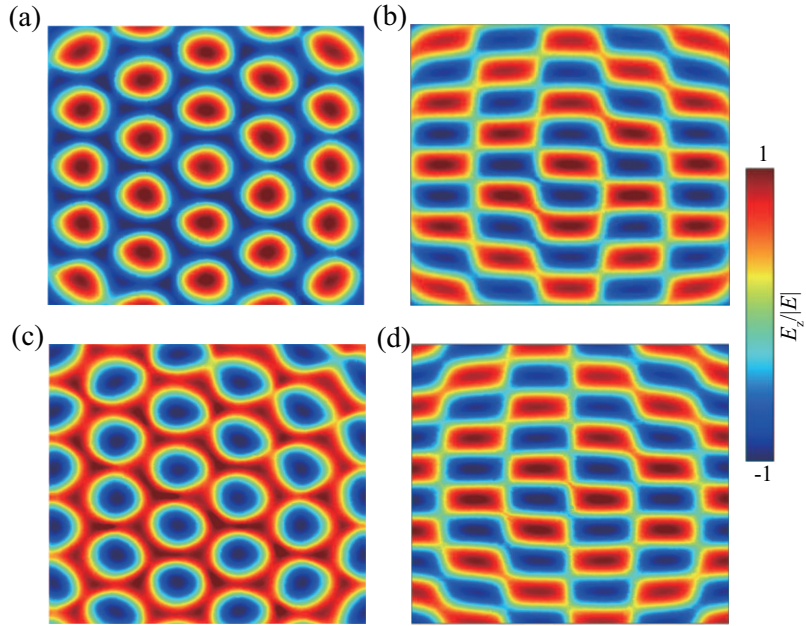}}
\caption{The intensity distribution of optical skyrmions electric field based on the full-wave electromagnetic simulations. (a) $E_{F14}=E_{F25}=E_{F36}=0.40~eV, \phi_{14}=\phi_{25}=\phi_{36}=0$. (b) $E_{F14}=0.20~eV, E_{F25}=E_{F36}=0.40~eV, \phi_{14}=\phi_{25}=\phi_{36}=0$. (c) $E_{F14}=E_{F25}=E_{F36}=0.40~eV, \phi_{25}=\pi, \phi_{14}=\phi_{36}=0$. (d) $E_{F14}=0.20~eV, E_{F25}=E_{F36}=0.40~eV, \phi_{25}=\pi, \phi_{14}=\phi_{36}=0$.
}
\label{fig4}
\end{figure}

To further understand the dynamic manipulation of optical skyrmions enabled based on the electrotunable properities of GPs and the tunable properties of the phase and polarization of the incident vector light field, full-wave electromagnetic simulations is adopted to solve the Maxwell equations by using the commercial software COMSOL Multiphysics based on finite element method (FEM). The configuration of the calculation is shown in Fig. \ref{fig1}(a), the single layer graphene is placed above SiO$_{2}$(300~nm)/Si substrate. Six slits of 1.5~$\mu$m length and 0.05~$\mu$m width were set in the graphene layer with the permittivity of 1. The distance between a pair of parallel graphene slit is 3.2~$\mu$m. Here, graphene is modeled as a thin film with the thickness of 0.34~nm and the in-plane conductivity of the graphene is computed within the local-random phase approximation (RPA)~\cite{falkovsky2007space32, gusynin2006unusual33}. The carrier mobility of the graphene is $\mu=8000~cm^{2}/Vs$~\cite{RN34} and ambient temperature is set as T=300~K. The permittivity of SiO$_{2}$ was taken from~\cite{palik1998handbook20}. In the simulation, the wavelength of electomagnetic field excitated by the circularly polarization light is $\lambda_{0}$=10.653~$\mu$m. The near-field signals were recorded as $E_{z}$ at the $x-y$ plane, 20~nm above the graphene. The optical skyrmions lattice is shown in Fig. \ref{fig4} by plotting the normalized axial electric field distribution at the center of the hexagon graphene slits structure with size of 3$\lambda_{GPs} \times$ 3$\lambda_{GPs}$ at 20~nm above the graphene surface.

Using the same circular polarization light to excite the graphene hexagonal slits with the Fermi energy of 0.40~eV, that is, $E_{F14}=E_{F25}=E_{F36}=$0.40~eV, $\phi_{14}=\phi_{25}=\phi_{36}$=0 and the optical skyrmions electric field distribution at the center of the structure as shown in Fig. \ref{fig4}(a). In the actual simulation, to compensate for the phase of circular polarization at different locations, the graphene hexagonal structure is moved outward by a distance of $n\cdot\pi/6$, $n=1, 2, 3, 4, 5$, respectively, which counterclockwise from the second slit. It can be known that, the result is consistent with the numerical simulation results obtained by Eq. (\ref{eqn2}) and Eq. (\ref{eqn3}) shown in Fig. \ref{fig3}(a). Moreover, the optical skyrmions topological states was also calculated when only changed the Fermi energy or the phase difference of the GPs standing wave. Fig. \ref{fig4}(b) and Fig. \ref{fig4}(c) corresponding to the optical skyrmions axial electric field distribution as only changed $E_{F14}$ from 0.40~eV to 0.20~eV, $E_{F25}=E_{F36}$=0.40~eV, $\phi_{14}=\phi_{25}=\theta_{36}$=0 and only changed $\theta_{14}$ from 0 to $\pi$, while $E_{F14}=E_{F25}=E_{F36}$=0.40~eV, $\theta_{25}=\theta_{36}$=0. In addition, changing the Fermi energy and the phase difference simultaneously, the distribution of the optical skyrmions axial electric field was shown in Fig. \ref{fig4}(d) with the three-dimensional electric field vector distribution at the bottom forming a N$\acute{\text{e}}$el-type skyrmion, where $\phi_{14}=\theta_{36}$=0, $\theta_{25}=\pi$, $E_{F25}=E_{F36}$=0.40~eV and $E_{F14}$=0.20~eV. By comparaing the distribution of optical skyrmion electric fleld before and after the change of Fermi energy and the phase difference, we can get the following conclusions: the shape of the optical skyrmions remains unchanged, but the direction of the vector electric field becomes opposite when only change the phase from 0 to $\pi$ of the GPs stangding wave; the shape of the optical skyrmions changes from a circle to a square when only change the Fermi energy from 0.40~eV to 0.20~eV of the GPs stangding wave; in addition, when the phase difference and the Fermi energy changes simultaneously, the shape and the direction of the vector electric filed of the optical skyrmions will change, which is equivalent to the supersition of the corresponding optical skyrmions distribution when change the Fermi energy and the phase difference independently. These results are consistent with the theoretical calculation above.

\section{Conclusion}
In summary, we proposed the dynamic regulation of optical skyrmions topological states continously based on the electrical tunable properties of graphene. The GPs skyrmions topological states is constructed by the interaction between incident vector light field and the graphene hexagonal structure. Therefore, the GPs skyrmions topological states can be regulated from two aspects of the vector light field and the graphene material, respectively. By adjusting the step of the Fermi energy changes of graphene, the continuous change of the topological states of the optical skyrmions can be achieved. Moreover, combining the optical regulation method based on vector light field with the electrical regulation method based on graphene materials can realize the mutual conversion and the continuous regulation between different topological states of the optical skyrmions, which greatly brodens the scope of regulation. Our work demonstrates that it is feasible to achieve the continous regulation of optical skyrmions based on the electrotunable properities of graphene and increases the degree of freedom of regulation, which provides a new idea for future integrated photonics devices.

\begin{backmatter}
\bmsection{Acknowledgments}
 This work was supported by National Natural Science Foundation of China (92050202, 12204309, 62275158); Shanghai Rising-Star  Program (22YF1415200).

\bmsection{Disclosures}
The authors declare no conflicts of interest.

\bmsection{Data Availability Statement}
Data underlying the results presented in this paper are not publicly available at this time but may be obtained from the authors upon reasonable request.

\end{backmatter}

\bibliography{ref1}

\end{document}